\documentclass[useAMS,usenatbib]{mn2e}

\usepackage{epsfig}
\usepackage{amsmath}
\usepackage{amssymb}
\usepackage{graphicx}
\usepackage{multirow}
\usepackage[font=small,labelfont=bf,labelsep=period]{caption}

\usepackage{comment}

\usepackage{color}

\title[Supernovae in structured and ionised clouds]
{The energy and momentum input of supernova explosions in structured and ionised molecular clouds}
\author[]
{Stefanie Walch$^{1}$ \& Thorsten Naab$^{2}$ \\
$^{1}$Physikalisches Institut, Universit\"at zu K\"oln,
Z\"ulpicher Strasse 77, 50937 K\"oln, Germany\\
$^{2}$Max-Planck-Institut f\"ur Astrophysik, Karl-Schwarzschild-Str. 1, 85741 Garching, Germany\\
}
\date{Accepted ???. Received ??? in original form ???}

\pagerange{\pageref{firstpage}--\pageref{lastpage}} \pubyear{2014}

\begin{document}
\label{firstpage}
\maketitle

\begin{abstract} 
We investigate the early impact of single and binary supernova (SN)
explosions on dense gas clouds with three-dimensional,
high-resolution, hydrodynamic simulations. The effect of cloud
structure, radiative cooling and ionising radiation from the
progenitor stars on the net input of kinetic energy, $f_{\mathrm{kin}}
= E_{\mathrm{kin}}/E_{\mathrm{SN}}$, thermal energy,
$f_{\mathrm{therm}} = E_{\mathrm{therm}}/E_{\mathrm{SN}}$, and gas
momentum, $f_{\mathrm{P}} = P/P_{\mathrm{SN}}$, into the interstellar
medium (ISM) is tested.  For clouds with $\bar{n} = 100\;{\rm
  cm}^{-3}$, the momentum generating Sedov and pressure-driven
snowplough phases are terminated early ($ \sim 0.01$ Myr) and
radiative cooling limits the coupling to $f_{\mathrm{therm}} \sim
0.01$, $f_{\mathrm{kin}} \sim 0.05$, and $f_{\mathrm{P}} \sim 9$,
significantly lower than without cooling. For pre-ionised
clouds the efficiencies are only increased by $\sim 50\%$, independent
of the cloud structure. This only suffices to accelerate $\sim$ 5\% of
the cloud to radial velocities $\gtrsim 30\;{\rm km\;s}^{-1}$. A
second SN might further enhance the coupling efficiencies if delayed
past the Sedov phase of the first explosion. Such very low coupling
efficiencies cast doubts on many galaxy-scale sub-resolution models
for supernova feedback, most of which are validated a posteriori
by qualitative agreement of galaxy properties with
observations. Ionising radiation appears not to significantly enhance
the immediate coupling of SNe to the surrounding gas as it drives the
ISM into inert dense shells and cold clumps, a process which is
unresolved in galaxy scale simulations. Our results support previous
conclusions that supernovae  might only drive a wind if a significant
fraction explodes in low-density environments or if they are supported
by processes other than ionising radiation.  
\end{abstract}


\section{Introduction}
\label{intro}
For plausible Galactic initial stellar mass-functions the minority of
newly formed stars are more massive than $\sim$ 8 solar masses
\citep{Chabrier2001}. However, with respect to their energy and
momentum output these stars are special. During their (short) lifetime
they emit highly energetic radiation, which ionises and heats the
surrounding cold filamentary and turbulent interstellar medium (ISM)
and changes its structure  by creating high density pillars and low
density cavities \citep{Gritschneder2009,Gritschneder2010,Walch2012b},
which may lead to triggered star formation \citep{Walch2013}. During a
lifetime of $\sim 5$ Myrs the total emitted energy of a typical O
star, which emits Lyman continuum photons at a rate of $\sim
10^{49}\,{\rm s}^{-1}$, amounts to more than $10^{52}$ erg. Although
only a small fraction ($<0.1$\%) of the emitted energy is transferred
to the ISM, ionising radiation seems to be sufficient to disperse low
mass ($\sim 10^4 {\rm M}_{\odot}$) molecular clouds (MCs) by
ionisation alone \citep{Walch2012b}. More massive, or very dense MCs
cannot be dispersed efficiently if their escape velocity is above the
velocity of the ionised gas $\sim 8\,{\rm km\; s}^{-1}$
\citep{2002ApJ...566..302M,Dale2012}.  
When a massive star explodes as a type II supernova (SN), an
additional amount of energy, $\sim 10^{51}$ erg, is almost
instantaneously released and drives a blast wave through the ambient
ISM. These supernova blast waves can disperse more massive MCs and
locally terminate the star formation process. A detailed understanding
of the coupling of SN energy to the structured multi-phase ambient ISM
is fundamental for several aspects of galaxy formation and evolution.  
 
It has been asserted early-on that type II supernovae might play a
crucial role for the evolution of galaxies in a cosmological context
\citep{Larson1974, Dekel1986, Navarro1993}. They can be an important
regulator of the galactic star formation rate \citep{Ostriker2010,
  Shetty2012, Hopkins2012}, eventually powering large scale galactic
winds \citep{Larson1974, Veilleux2005, Joung2006, Dubois2008,
  Creasey2013}, which might drive the enrichment of the ISM with heavy
elements. However, in most simulations of galaxy formation the
relevant temporal and spatial scales within MCs are unresolved and the
supernova energy is typically released on the minimum resolution scale
($\sim 10^{4}\,{\rm M}_{\odot}$ and $\sim 50\,{\rm pc}$ for high
resolution cases). This procedure leads to a diversity of problems
depending on the resepective implementation. A wealth of 'plausible' 
sub-resolution models have been developed. However, most of them
cannot of do not follow the temperature structure of the ISM below
temperatures of $10^4\,{\rm K}$ and assume the transformation of gas
to stars below a relatively 
low density threshold ($n \sim 0.1 - 1\, {\rm cm}^{-3}$)
representative for the ionised gas phase in the ISM \citep{Schaye2014, 
  Vogelsberger2014}. In this paper, we 'zoom in' on one of the 
cosmological resolution elements and aim to understand the SN feedback 
within. 

Analytically, SNe have been well studied in early similarity solutions
by \citet{Sedov1959} and \citet{Taylor1950}. In a homogeneous medium a
spherical Sedov-Taylor (ST) blast wave (at constant total energy)
develops after an initial phase of free expansion of supernova ejecta
(at constant velocity) during which the shock sweeps up the
surrounding gas. Early calculations indicate that the swept-up mass
can reach $\sim 30$ times the ejecta mass before the reverse shock
reaches the center of the explosion and the ST phase begins
\citep{Gull1973}. Both, the free expansion and the ST phase are termed
      {\it non-radiative} as energy losses by radiation are
      dynamically insignificant (see \citet{Truelove1999} for a
      unified solution for both processes).  

Energy losses by radiation become important once the cooling time of
the post-shock material becomes shorter than the flow time and the
system enters first a pressure-driven and then a momentum-conserving
snowplough phase 
\citep{Ostriker1988, Cioffi1988}. Early analytical \citep{McKee1977}
and numerical \citep{Cowie1981} estimates have found the energy
conversion efficiency, i.e. the fraction of the explosion energy which
is finally retained in the ambient medium, to be $\sim$50\%. This
amount would explain the observed thermal pressure in the ISM of the
Milky Way by SN energy input alone.  

However, simulations by \citet{Cowie1981} indicate a strong dependence
of the energy conversion efficiency on the ambient gas density. For
number densities $> 1 \,{\rm cm}^{-1}$, radiative losses become
increasingly important and eventually result in a small remaining
fraction of hot gas at $\gtrsim 10^5\; {\rm yr}$ post
explosion \citep{Slavin1992}. In this case, only a few percent of the
initial SN energy might be retained in the ISM. The bottom line is,
that the properties of the environment within which a supernova
explodes strongly affect the fraction of energy that is deposited into
the ISM \citep{Dwarkadas2012}.  

For this reason, it is essential to consider the structure of the ISM
that surrounds the massive progenitor before the SN explodes. In
particular, the vicinity of the SN progenitor is shaped by ionising
radiation and stellar winds. In one- and two dimensional simulations
\citet{Dwarkadas2005} and \citet{Dwarkadas2007} have investigated SN explosions in
wind-blown bubbles. They show that a wind can significantly disturb
the initially homogeneous ISM before the SN explodes. In such an
environment the ST phase is reached later than expected for a
homogeneous medium of the same density. Alternatively, the explosion
might not develop a ST phase at all, if radiative cooling is
efficient. 

A supernova remnant evolving in this highly
structured environment is expected to behave differently with respect
to the well studied explosions in homogeneous media.
\citet{Rogers2013} have simulated the evolution of very low mass 
(few $\times 10^3\,{\rm M}_{\odot}$) turbulent clouds under the influence of 
stellar winds and SN explosions. They find that the SN energy is only weakly coupled
to the ambient ISM since the winds have already dispersed the clump significantly. 
In addition to winds, ionizing radiation heats the ambient
medium and shapes the density structure of the molecular cloud (MC),
thereby creating high density pillars and low density cavities
(e.g. \citealp{Dale2009, Gritschneder2010, Peters2010,
  Walch2012b}). Ionisation therefore broadens the density probability
density distribution (PDF) of a MC and can initiate outflows \citep{Walch2012b}.  However,
detailed simulations of SNe exploding within ionised bubbles are still missing.

In this paper we present high-resolution, three-dimensional
simulations of a single or binary supernova explosions within a representative MC of
$\sim 10^5 {\rm M}_{\odot}$. The MC may have a fractal substructure
and has been pre-ionized by the massive progenitor star. By comparing
simulations in structured and homogeneous MCs, as well as with and
without metal-cooling, we are able to quantitatively derive the momentum input
gained by the SN explosions, a quantity which is highly valuable for large-scale
simulations that cannot resolve the ST phase of the SNe.

The paper is organized as follows. In section \ref{sec2} we describe the SPH code \textsc{Seren} including the treatment of ionizing radiation, SN explosions and radiative cooling. In section \ref{sec3} we show Supernova explosions in homogeneous and fractal clouds. The energy and momentum input is discussed in section \ref{SEC_ENERGETICS}, and in \ref{SEC_OUTFLOW} we show the evolution of mass, energy, and momentum in different temperature regimes. Our main conclusions are summarized in section \ref{Conclusions}.

\section{Numerical method and initial conditions}\label{sec2}
\subsection{SPH code \textsc{Seren} with ionization}\label{SEC_SEREN}
We use the SPH code \textsc{Seren} \citep{Hubber2011}, which is well-tested and has been applied to many problems in star formation \citep[e.g.][]{Bisbas2011, Walch2012a, Walch2013, Lomax2014}. We employ the standard SPH algorithm \citep{Monaghan1992}. The SPH equations of motion are solved with a second-order Leapfrog integrator, in conjunction with a block time-stepping scheme. Self-gravity of the gas is included and the gravitational forces are calculated using an octal-spatial decomposition tree \citep{Barnes1986}, with monopole and quadrupole terms. We use a multipole acceptance (tree-opening angle) criterion that controls the relative gravitational acceleration error \citep{Springel2001} and a standard artificial viscosity prescription \citep{Monaghan1983}.
{\sc Seren} has been demonstrated to perform well in standard hydrodynamic test problems \citep{Hubber2013}.

In runs with ionization a full radiation- hydrodynamical calculation is performed. The ionizing radiation is treated with a HEALPix-based adaptive ray-splitting algorithm, which allows for optimal resolution of the ionization front to twice the SPH particle smoothing length \citep[see][]{Bisbas2009}. Our treatment of ionization and thermal balance is simplistic. We compute hydrogen ionization and adopt the On-The-Spot Approximation \citep{Osterbrock2006} to treat the diffuse component of the Lyman Continuum radiation. Thus, shadowed regions cannot be ionized. The ionized gas is heated instantaneously to a temperature of $10^4\,{\rm K}$; in other words, we neglect variations in the temperature due to -- for example -- the hardening of ionizing photons with increasing distance from the ionizing star. The neutral gas is assumed to be at a temperature of $30\,{\rm K}$, which is the initial temperature of the simulated MC (see section \ref{SEC_IC}). We assume an O-star progenitor that emits $10^{49}\;{\rm s}^{-1}$ Lyman continuum photons per second before it explodes as a type II SN. Once the SN explodes, the ionising radiation is switched off. 

\subsection{Radiative cooling }\label{SEC_COOL}
{\sc Radiative cooling:} We assume solar metallicity for all simulations and include radiative cooling in two different regimes. (1) For gas with temperatures $T>10^4$ K we compute the cooling rate $\Lambda_{_{\rm cool}}$ for every SPH particle from interpolating the cooling table by \citet{Plewa1995}, which is a function of density and temperature. (2) For $T\le 10^4$ K, we derive $\Lambda_{_{\rm cool}}$ from the analytical formula by \citet[][hereafter KI]{KI2000, KI2002}:
\begin{eqnarray} \label{EQ_KI}
 \Lambda_{_{\rm cool}} & =& \Gamma  \left[10^7 {\rm exp}\left(\frac{-1.184 \times 10^5}{T + 1000} \right) \right. \label{EQ_KI}\\
& &+ \left. 1.4 \times 10^{-2} \sqrt{T}\; {\rm exp}\left(\frac{-92}{T} \right)\right] \frac{\rm erg\; cm^3}{\rm s} \nonumber
\end{eqnarray}
with the fixed heating rate $\Gamma$ of
\begin{eqnarray}
\Gamma &= & 2.0 \times 10^{-26} \; \frac{\rm erg}{\rm s} \nonumber.
\end{eqnarray}
Eq.\ref{EQ_KI} has been derived from highly resolved, one-dimensional radiation- hydrodynamical simulations of collapsing MCs at solar metallicity.
For both temperature regimes we compute the change in specific internal energy, $\dot{u}$, of each SPH particle within the local SPH timestep $\delta t$ according to
\begin{equation}
 \dot{u}= -\Lambda_{_{\rm cool}} \times n^2 / \rho \times \delta t, 
\end{equation}
where $\rho$ is the mass density and $n$ the number density of the particle. 
We do not allow the gas to cool below a minimum temperature of 30 K and restrict $\dot{u}$ accordingly.

If the local cooling time, $\tau_{_{\rm cool}}$, is shorter than $\delta t$, we sub-cycle the calculation of $\Lambda_{_{\rm cool}}$ and $\dot{u}$. The local cooling time is defined as 
\begin{equation}\label{EQ_TCOOL}
 \tau_{_{\rm cool}}= \frac{u \rho}{\Lambda_{_{\rm cool}} n^2}.
\end{equation}
If $\tau_{_{\rm cool}} < \delta t$, $\delta t$ is divided into $n$ substeps, where $n=\delta t/\tau_{_{\rm cool}}$. In this way the temperature update is more accurate and artificial over-cooling or -heating within one $\delta t$ is reduced.

\subsection{ Supernova explosions} \label{SEC_SN}

To model a Supernova explosion we add $8\; {\rm M}_\odot$ of ejecta mass in a spherical 'injection region' with a radius of $0.1$ pc around the explosion center. The ejecta particles have the same mass as the ambient ISM particles and therefore the SN ejecta are resolved with $\sim$ 80 SPH particles (see section \ref{SEC_IC}). Every ejecta particle is given a radial velocity of approximately $3400\,{\rm km \,s}^{-1}$ -- approximately, because the ISM particles that are present within the injection region are also accelerated -- corresponding to a a total energy of $E_0=10^{51}$ erg \citep[see e.g.][]{Janka2013} and a radial momentum of $P_0 = 2.77\times 10^{4}\,{\rm M}_\odot\,{\rm km \,s}^{-1}$ ($P_0$ is sufficient to accelerate about one third of the MC to $\sim 1\,{\rm km\,s}^{-1}$). Since we will also study the case of binary explosions, we denote the total energy and momentum added to the simulation with $E_{_{\rm SN}}$ and $P_{_{\rm SN}}$. 

Although the momentum and energy input in the form of adding some high-velocity ejecta mass is probably the most realistic implementation, we have also tested the injection of $E_\mathrm{SN}$ in the form of thermal energy.
In the adiabatic case, we find that both injection methods lead to an equally good reproduction of the ST solution if we strictly and conservatively reduce the individual time steps of all particles in the vicinity of the SN explosion (at the very least $\delta t$ of all neighbours to the 'explosion particles' must be reduced). This confirms the results of \citet{Hubber2011}, who first introduce this time step criterion \citep[see also][]{Durier2012}.

\subsection{Initial conditions}\label{SEC_IC}
                                                         
In this study, we simulate a typical molecular cloud with a total mass of ${\rm M}_{_{\rm MC}}=10^5\; {\rm M}_\odot$, and a radius of $R_{_{\rm MC}} =16 \;\rm{pc}$. The mean density is $\bar{\rho}_0 = 3.94 \times 10^{-22} \rm{g \;cm}^{-3}$, or equivalently $\bar{n} \approx 100 \;\rm{cm}^{-3}$ for molecular gas with a mean molecular weight $\mu=2.35$. These conditions resemble a single, or a small number of SPH particle(s) in simulations of galaxy formation \citep[e.g.][]{Naab2007}.

The escape velocity of this cloud is $v_\mathrm{esc}\approx \sqrt{2 G {\rm M}_{_{\rm MC}}R^{-1}} = 7.3\; {\rm km\; s}^{-1}$, and in the case of a homogeneous cloud we derive a gravitational binding energy of $E_\mathrm{pot}=\frac{3}{5} \frac{G M^2}{R} \approx 3 \times 10^{49} {\rm erg}$. Initially, the cloud gas is isothermal and cold $T=30$ K. The initial thermal energy of the cold gas, $E_\mathrm{therm,0} = 2.1 \times 10^{47}\; {\rm erg}$, is small compared to the SN energy input.

We use $10^6$ SPH particles, thus each particle has a mass of $m_{_{\rm part}}= 10^{-1}\;{\rm M}_\odot$. For simulations with a uniform density distribution we construct a glass of SPH particles. 
In the following we describe the different simulation setups. Their key parameters and settings are also summarised in Table \ref{TABLE_1}.

\begin{table*}
\begin{tabular}{lccccccclcc}
Run & $\bar{n}$ & Structure & Gas physics & Binary & $t_2$ & $E_\mathrm{SN}$ & $P_\mathrm{SN}$ & $\;\;\;\;f_{_{\rm P}}$ & $f_\mathrm{kin}$ & $f_\mathrm{therm}$\\

 & [$\frac{1}{\rm cm^{3}}$]&  & & & [kyr] & [10$^{51}$ erg] & [$10^4 \,\frac{\rm M_{\odot} \,{\rm km}}{\rm s}$] & $=\frac{P}{P_\mathrm{SN}}$ & $=\frac{E_\mathrm{kin}}{E_\mathrm{SN}}$ & $=\frac{E_\mathrm{therm}}{E_\mathrm{SN}}$\\
  \hline \hline
HA    & 100 & homogeneous & adiabatic  & 0 & 0 & 1 & 2.77 & 51.6 (+)  & 0.28 & 0.72 \\
HC   & 100  & homogeneous & cooling  & 0 & 0 & 1 & 2.77 &9.64 & 0.04 & 0.01 \\
HCI  & 100  & homogeneous & ionisation + cooling & 0 & 0 & 1 & 2.77 & 14.8 & 0.08 & 0.12 \\
FA    & 100  & fractal & adiabatic & 0 & 0 & 1 & 2.77 & 49.1 (+) & 0.29 & 0.57 \\
FC   & 100  & fractal & cooling & 0 & 0 & 1 & 2.77 & 8.90 & 0.05 & 0.01\\
FCI  & 100  & fractal & ionisation + cooling & 0 & 0 & 1 & 2.77 & 14.0 & 0.09 & 0.14 \\
FCN10 & 10  & fractal & cooling & 0 & 0 & 1 & 2.77 & 12.2 & 0.11 & 0.03\\
FCN1 & 1 & fractal & cooling & 0 & 0 & 1 & 2.77 & 14.4 (+) & 0.23 & 0.11\\
FCB0 & 100  & fractal & cooling & 1 & 0 & 2 & 5.54 & 8.40 & 0.05 & 0.01\\
FCB5 & 100  & fractal & cooling & 1 & 5 & 2 & 5.54 & 8.76 & 0.05 & 0.04 \\
FCB20 & 100  & fractal & cooling & 1 & 20 & 2 & 5.54 & 11.4 & 0.08 & 0.12 \\
FCB100 & 100  & fractal & cooling & 1 & 100 & 2 & 5.54 & 9.95 (+)& 0.08 & 0.29\\
FCIB100 & 100  & fractal & ionisation + cooling & 1 & 100 & 2 & 5.54 & 11.5 (+) & 0.09 & 0.31\\
\end{tabular}
\caption{List of all simulations. Column 1 gives the simulation identifier; column 2 gives the mean number density of the cloud; column 3 gives the density structure, where we use the same particular setup for all runs with fractal or homogeneous structure, respectively. Column 4 describes the gas cooling and ionisation state of the gas cloud. Column 5 specifies the binary properties, i.e. if $b=0$ we have a single explosion and if $b=1$ we have two explosions. The time of the second explosion, $t_2$, is listed in column 6 (the first explosion always takes place at $t=0$). In column 7 and 8 we give the energy and momentum injected by the SN explosions. In column 9, 10, and 11 we list the ratio of the current gas momentum to the input momentum, $f_\mathrm{P}$, the fraction of retained kinetic energy, $f_\mathrm{kin}$, and the fraction of retained thermal energy, $f_\mathrm{therm}$, at $t=0.2$ Myr.}\label{TABLE_1}
\end{table*}


\subsection{Simulation setup}\label{SEC_IC}

\subsubsection{ Homogeneous molecular clouds} 
In the first set of simulations, we consider homogeneous MCs, that have the same mean density and total mass. The SN is always initiated in the centre of the cloud at $t=0$. We carry out three types of simulations: 
\begin{enumerate}
\item[1.] Run HA: Homogeneous \& Adiabatic. The central SN explosion is evolved with an adiabatic equation of state with adiabatic index $\gamma=\frac{5}{3}$, i.e. the conditions for a Sedov-Taylor explosion \citep{Sedov1959, Taylor1950}. 
\item[2.] Run HC: Homogeneous \& metal-line Cooling: Same setup as in HA but with additional radiative cooling as described in section \ref{SEC_COOL}
\item[3.] Run HCI: Homogeneous \& metal-line Cooling \& Ionisation: Here, we take into account that the massive star progenitor has emitted ionising radiation prior to the SN. To model the initial HII region expansion, we place a source of ionising radiation emitting $\dot{N}_{_{\rm LyC}}=10^{49}\, {\rm s}^{-1}$ at the center of the MC and evolve it for 1 Myr \citep[see][]{Walch2012b}. This is done with the Healpix-based ionisation scheme as described in section \ref{SEC_SEREN}. The MC is much larger than the Str\o mgren radius as well as the evolved H{\sc ii} region at the explosion time. Therefore no outflow is initiated by the expansion of the H{\sc ii} region. The subsequent SN is evolved with radiative cooling as in run HC.
\end{enumerate}

\subsubsection{Fractal molecular clouds} 
In a second set of simulations we investigate the influence of the cloud sub-structure. Therefore, we initialize the MC with a self-similar, fractal density distribution, which is frequently observed in turbulent, cold molecular clouds \citep[e.g.][]{Stutzki1998, Sanchez2005}.
Following \citet{Walch2012b} the fractal clouds are constructed by setting up a density power spectrum in Fourier space. The power spectral index $n$ is directly related to the fractal dimension of the density field $\mathcal{D}$:
\begin{equation}
\mathcal{D} = 3- \frac{\left(n-2 \right)}{2}
\end{equation}
The density field is then scaled to give a log-normal density PDF and the same mean density of $\bar{n} \approx 100\;{\rm cm}^{-3}$ as the homogeneous cloud \citep[see also][]{Shadmehri2011}. The MC has $D=2.6$, since we have previously shown that this fractal dimension is prone to lead to the development of particularly realistic HII regions \citep{Walch2012b, Walch2013}. 
Again, we run three simulations for the fractal MC (see Table \ref{TABLE_1}):
\begin{enumerate}
 \item[1.] Run FA: Fractal \& Adiabatic. The SN is placed in the center of the cold, fractal molecular cloud and is evolved with an adiabatic equation of state.
\item[2.] Run FC: Fractal \& metal-line Cooling: Same setup as in FA plus additional radiative cooling.
\item[3.] Run FCI: Fractal \& metal-line Cooling \& Ionisation: The SN explosion is evolved with radiative cooling in a fractal cloud which is pre-ionised by the SN progenitor. To simulate the initial HII region we follow the same procedure as for run HCI.
\end{enumerate}

\subsubsection{Low density molecular clouds}
 Since the energy lost due to radiative cooling is proportional to $\bar{n}^2$, the MC density could be a key parameter for determining the evolution of the SN remnant \citep[e.g.][]{Cioffi1980}. We test this possibility in two simulations with a reduced mean density of $n=\,10\,{\rm cm}^{-3}$ (run FCN10) and $n= 1\,{\rm cm}^{-3}$ (run FCN1). We use the same fractal substructure and the same cloud mass, i.e. the MCs are increased in size to give the required $\bar{n}$. Hence, the new outer radii are $R_{_{\rm MC}} = 34.5\,{\rm pc}$ for FCN10 and $R_{_{\rm MC}} = 74.3\,{\rm pc}$ for FCN1.

\subsubsection{Binary explosions}
Most massive stars live in binary systems or small number associations \citep[see][and references therein]{Reipurth2014}. It is hence likely that a second SN explodes within the bubble that is created by the explosion of the primary star, thus creating a super-bubble \citep{Oey1997, Preibisch2007, Wunsch2011, Higdon2013}. We investigate the impact of binary explosions that take place with different offsets in time, $t_2$. We choose $t_2$ such that the second explosion is set off in different phases of the first SN remnant. In particular, we consider $t_2=5\,{\rm kyr}$ (run FCB5), when the first SN remnant is still in the ST phase; $t_2=20\,{\rm kyr}$ (run FCB20), when the first SN remnant has just entered the pressure-driven snowplough phase; and $t_2=100\,{\rm kyr}$ (run FCB100), when the first SN remnant is well into the snowplough phase. The simulations are compared to the case where we inject twice the amount of SN energy at once, i.e. $2\times 10^{51}\,{\rm erg} = 2 E_0$ are injected at $t=0$, and thus $t_2 = 0\,{\rm kyr}$ (run FCB0).


\begin{figure}
 \includegraphics[width=90mm]{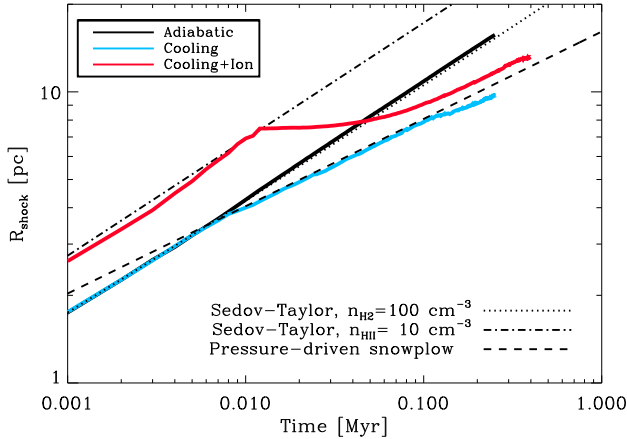}
 \caption{Time evolution of the shock front radii computed for runs HA (black solid line), HC (blue line), and HCI (red line). We also plot the ST solution, which fulfills $R\propto t^{2/5}$, for two different number densities; $n=100\;{\rm cm}^{-3}$ for molecular gas with $\mu=2.35$ (dotted line) and $n=10\;{\rm cm}^{-3}$ for ionised gas with $\mu=0.68$ (dash-dotted line). In the radiative stage, the expansion follows $R\propto t^{2/7}$, which is typical for the pressure-dominated snowplow phase (dashed line). For run HC the transition between ST phase and radiative phase occurs when about 50\% of the injected energy is radiated away. }\label{FIG_IFPOS}
\end{figure}


\section{Results} \label{sec3}

\subsection{SN remnant evolution in homogeneous clouds} 
In the homogenous case, the SN bubble expands in spherical symmetry, modulo very small fluctuations
that are caused by the random distribution of the injected SN particles. To compare our results to the ST solution for the three different cases (HA, HC, and HCI), we follow the propagation of the SN shock front radius, $R_\mathrm{shock}$, which is derived by averaging the radial location of the density maximum over 48 bins in azimuth. This criterion does not work well for the shock expansion within the HII region (run HCI) since, in this case, the density maximum is initially located within the cold swept-up shell and does not trace the SN shock. Therefore, we use an additional temperature criterion to identify $R_\mathrm{shock}$ in the early stages of HCI, i.e. only particles with $T> 10^5$ K are considered. In Fig. \ref{FIG_IFPOS} we plot $R_\mathrm{shock}$ as a function of time. The curves end at time $t_{_{\rm OUT}}^0 = 0.2 - 0.3\,{\rm Myr}$, at which the outflow of gas through the cloud surface sets in. For $t>t_{_{\rm OUT}}^0$ the definition of $R_\mathrm{shock}$ is not reliable anymore.

In the simplest case of an adiabatic explosion in a homogeneous medium (run HA) the ST solution is recovered. This means that the expansion of the SN shock follows a powerlaw:
\begin{equation}
R_\mathrm{shock, ST}(t) = \beta^{-1} \left(\frac{E_0}{\bar{\rho}_0}\right)^{1/5} t^{2/5},
\end{equation}
with $\beta=0.868$ for adiabatic index $\gamma=\frac{5}{3}$ \citep{Ostriker1988}. 
We overplot the ST solution for two particular densities in Fig. \ref{FIG_IFPOS}. For the initial number density of molecular gas $\bar{n}=100\;{\rm cm}^{-3}$ (with $\mu=2.35$; dotted line) and for the initial number density of ionised gas within the HII region $\bar{n}=10\;{\rm cm}^{-3}$ (with $\mu=0.68$; dash-dotted line). 
 
The ST solution is only applicable (after the swept-up mass is greater than the ejecta mass and) as long as the total energy is constant. When radiative cooling becomes important this condition is violated and the time dependence of the shock propagation changes. This phase is called the radiative snowplough phase. To be precise, there are two sequent radiative snowplough phases: The pressure-driven snowplough (PDS) phase, in which the rarefied medium within the bubble is still hot enough to provide significant pressure, and the momentum-conserving snowplough. Here, we only describe the PDS stage, which contributes to the momentum input. In the PDS stage the shock front expansion follows:
\begin{equation}
R_\mathrm{shock, PDS}(t) = R_c \left(\zeta \frac{E_0}{\bar{\rho}_c R^5_c}\right)^{1/7} t^{2/7},
\end{equation}
with $\eta = 2/7$ the power law index of the expansion (see \citet{Ostriker1988}, their eq. (6.14)), and $\zeta= \frac{(\eta/2)^2}{4 \pi}$. $R_c$ is the radius at which half the energy of the initial blast wave has been radiated away, and $\bar{\rho}_c$ is the mean density of the cloud at this point (see dashed line in Fig. \ref{FIG_IFPOS}).\\

In run HC with radiative cooling, we find a transition from the ST phase to the PDS phase at $t\approx 8$ kyr. This is somewhat later than the transition time we derive following  \citet{Blondin1998}, $t \approx 2.1\,{\rm kyr}$, who assume a cooling rate of the form $\Lambda(T) \approx 10^{-16} T^{-1} \,{\rm erg\, cm}^{-3}\,{\rm s}^{-1}$ and a mean molecular weight of 0.62. This is probably due to the fact that their assumed functional form of the cooling rate is only valid within a limited temperature regime of $3\times 10^{5} \lesssim T \lesssim 10^7\,{\rm K}$, where the cooling rate is overall at least a factor of 10 higher than at $T \lesssim 3\times 10^{5} \,{\rm K}$.

\begin{figure*}
$\;\;\;\;\;\;\;\;\;\;\;\;\;\;\;\;\;\;\;\;\;\;\;\;\;\;\;\;\;\;\;\;\;$ {\bf log (Column Density /$[{\rm M}_{\odot} \;{\rm pc}^{-2}]$) }$\;\;\;\;\;\;\;\;\;\;\;\;\;\;\;\;\;\;\;\;\;\;\;\;\;\;\;\;\;\;\;\;\;\;\;\;\;\;\;\;\;\;\;\;\;\;\;\;\;${\bf  log (Temperature/[K])}\\
\includegraphics[width=170mm, trim= 0 2cm 0 0]{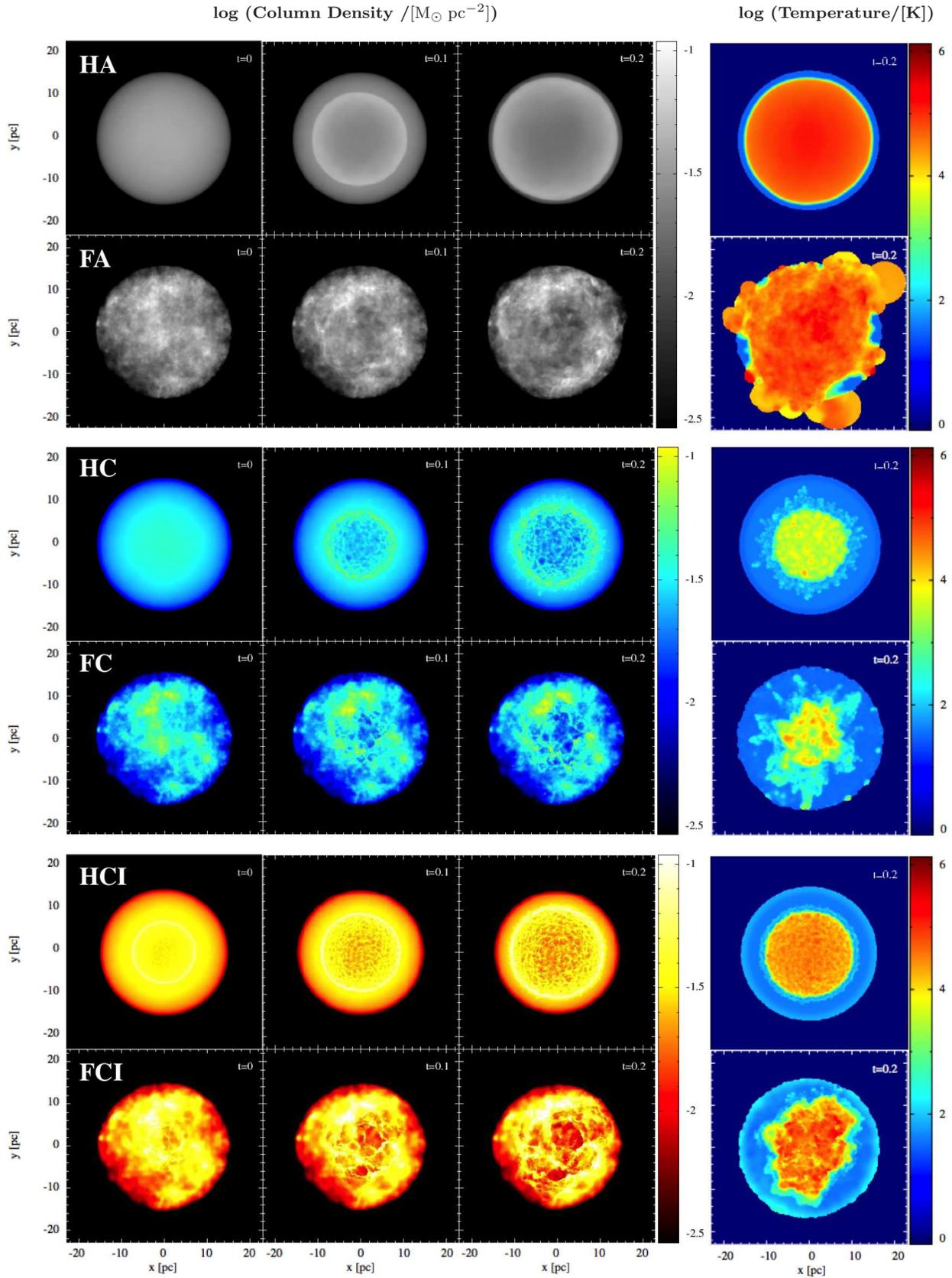}
\caption{{\it Left to right:} The first three columns show the column density (projected along the $z-$axis) at times $t=0.0$ and $t= 0.1$, and 0.2 Myr after the SN explosion; the fourth panel shows the maximum temperature along the line of sight for all six simulations. {\it Top to bottom:} We pair homogeneous and fractal clouds into groups of the same underlying gas physics: HA and FA (adiabatic), HC and FC (cooling), and HI and FI (ionisation and cooling). Both, the structure of the clouds and the ionising radiation modulate the impact of the SN shell. \label{FIG_CD}}
\end{figure*}

In run HCI the SN explodes within the initial HII region, where the gas has a mean number density of $\sim 10 \;{\rm cm^{-3}}$ and a temperature of 10,000 K. The expansion of the HII region has already swept up a shell of dense neutral gas within the MC (see panel showing $t=0$ in Fig. \ref{FIG_CD}). This shell is much more massive than the SN ejecta mass. As a result, the initial expansion of the SN follows the ST solution but becomes dominated by radiative cooling once the shock hits the dense swept-up shell, i.e. at $t\approx 20$ kyr. At this point, $R_\mathrm{shock}(t)$ stalls until enough momentum has been transferred to the shell to move it further outwards $t\approx60 $ kyr. From this time onward the remnant is in the snowplough phase.


\begin{figure*}
\includegraphics[width=88mm]{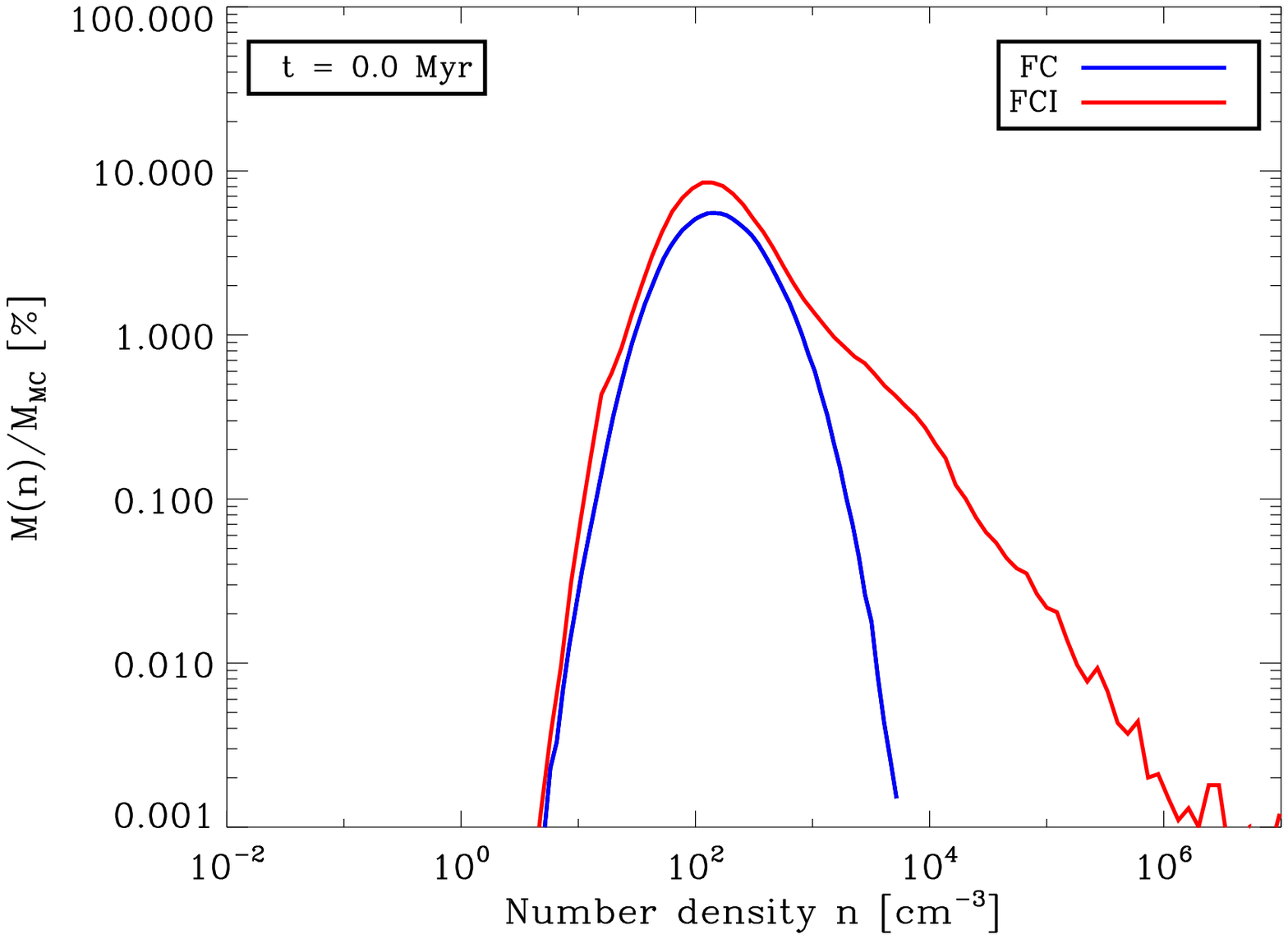} 
\includegraphics[width=88mm]{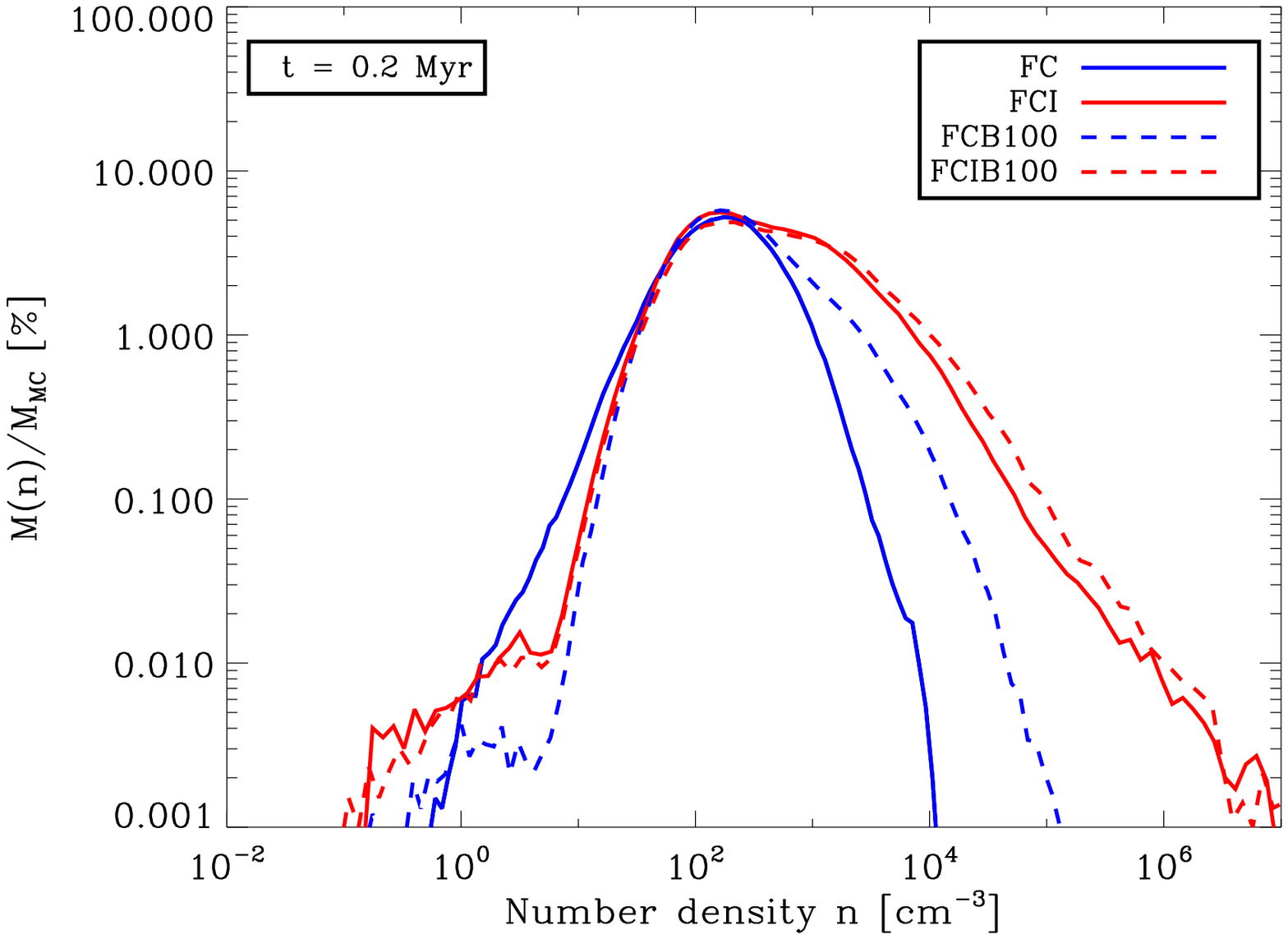} \\
\caption{{\it Left panel:} Density PDFs of the initial conditions for runs FC (same as FA; blue line), and FCI (red line). The initial condition for the fractal cloud is a log-normal density PDF. Ionising radiation results in a broader distribution, which extends to higher densities. The initial conditions of all homogeneous runs (not shown here) are given by a delta function at $\bar{n}=100\,{\rm cm}^{-3}$. 
{\it Right panel:} Density PDFs for these simulations at $t=0.2\,{\rm Myr}$. In addition, we show the PDFs for runs FCB100 (blue dashed line) and FCIB100 (red dashed line). \label{FIG_rho_PDF}}
\end{figure*}


\begin{figure*}
\includegraphics[width=150mm]{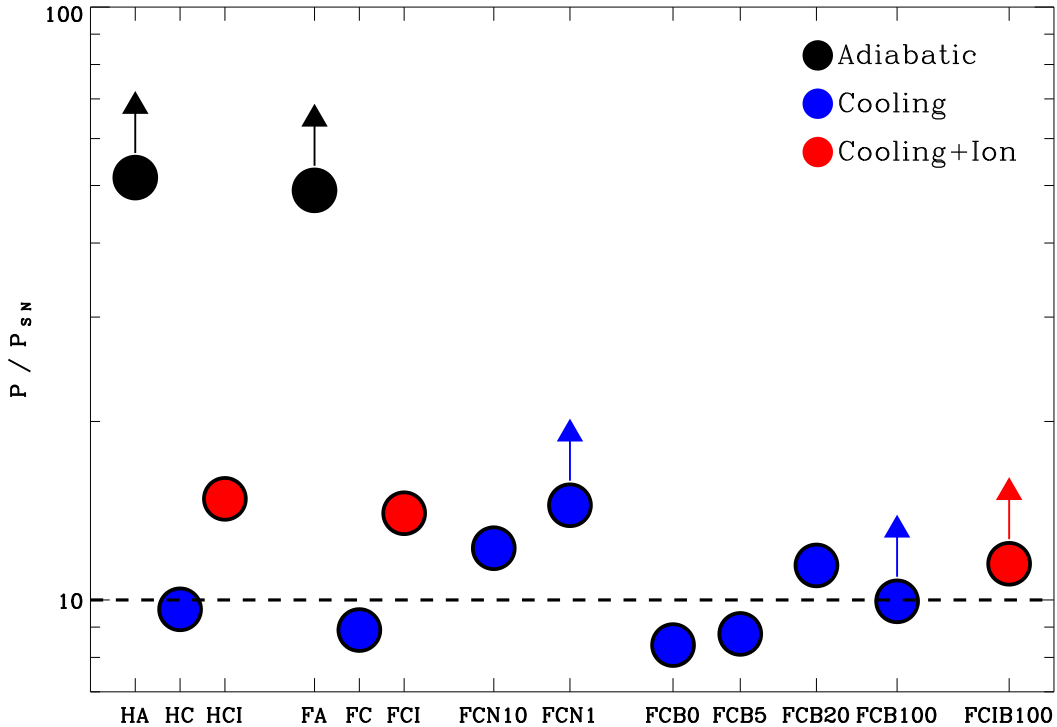}
\caption{Gas momentum injected into the ISM at $t=0.2$ Myr relative to
  the initial momentum of the supernova ejecta, $P_\mathrm{SN}$ for all runs. For a single explosion $P_\mathrm{SN} = 2.77 \times 10^4\; {\rm M}_\odot {\rm km/s} $, i.e. for all binary cases $P_\mathrm{SN} = 5.53 \times 10^4\; {\rm M}_\odot {\rm km/s}$. Cooling reduces the momentum input by more than a factor of 5. Ionisation only increases the momentum input by $\sim$ 50\% with respect to cooling. The cloud structure is of minor importance (e.g. HCI vs. FCI) and binary explosions have to be separated by more than 20 kyr to have a noticeable effect. \label{FIG_EPMAX}}
\end{figure*}


\subsection{SN remnant evolution in structured clouds}\label{SEC_STRUCTURE}

In Figure \ref{FIG_CD}, we show the time evolution of the column densities for six simulations at $t=0.0,  0.1,\,{\rm and}\, 0.2$ Myr. The right-most column shows the temperature distribution at $t=0.2$ Myr. We group the simulations according to the used gas physics. The adiabatic runs HA and FA are colour-coded in grey-scale (top panels), followed by runs HC and FC in blue (middle panels), which include radiative cooling. Runs HCI and FCI with ionisation and radiative cooling are colour-coded in red (bottom panels). The adiabatic simulations serve as a reference as they are energy conserving. Therefore, the SN explosion has the highest possible impact on the surrounding and the MCs are fully disrupted on timescales $t < 1$ Myr (depending on the setup, gas outflow starts at $t\approx 0.2$ Myr). In runs with radiative cooling, the bubble is much cooler and smaller, highlighting the importance of including cooling in the early phases of SN explosions in dense MCs. The swept-up shell also looks perforated due to cooling instabilities. If the SN explodes in an evolved HII region, the SN remnant remains hotter and the shell radius is larger than in runs HC and FC.

From Fig. \ref{FIG_CD} we can qualitatively assess that the impact of the SN explosion is sensitive to both, the underlying gas physics as well as the initial structure of the MC. In a fractal molecular cloud, the SN shell expands more slowly along directions where the column density, as seen from the explosion centre, is high. On the other hand, the SN shock wave can progress faster where the column density is low. 
The underlying fractal cloud structure leads to a complex shell geometry and allows for an early outflow of gas through low density regions, which evolve into holes in the cloud.
Due to the quick leakage of hot gas from the fractal cloud, the cloud vicinity is rapidly affected by the SN remnant. 

The effect of the expanding H{\sc ii} region prior to the SN explosion, run FCI, is twofold: (i) the gas surrounding the SN is heated to $10^4\,{\rm K}$ and (2) the initial density distribution within the cloud is changed by the ionising radiation. Instead of sweeping up a dense shell (as in run HCI) the H{\sc ii} region in FCI has a complex structure. It is bordered by compressed, dense pillars and shell-like structures as well as a significant fraction of low-density regions \citep{Walch2012b}. To demonstrate how the expansion of the H{\sc ii} region affects the MC, we show the density PDFs of the pristine fractal cloud (initial condition of FA and FC) and of the ionised fractal cloud (initial condition of FCI) in Fig. \ref{FIG_rho_PDF} (left panel). The ionised cloud features a broader density distribution, which mostly extends to higher densities. Therefore, the ionised configuration can be adjuvant for the SN efficiency in radial directions of low column density where radiative cooling is significantly reduced. At the same time, however, radiative cooling is significantly increased whenever the remnant hits a dense shell or pillar, where it immediately transits into the radiative stage. Thus, we expect the overall efficiency of the SN to be moderately enhanced if it explodes within an H{\sc ii} region (see section \ref{SEC_ENERGETICS}). \citet{Walch2012b} have discussed the surface coverage fraction, i.e. the fraction of the sphere's surface where the ionisation front is bound by dense gas, and have shown that it is independent of the fractal dimension of the cloud, for $\cal{D}$ ranging from 2.0 to 2.8. Thus we may assume that our findings are also applicable to clouds with different $\cal{D}$.

In the right panel of Fig. \ref{FIG_rho_PDF}, we show the density PDFs of runs FC, FCI, FCB100, and FCIB100 at $t=0.2$ Myr. Similar to the effect of ionisation feedback, Supernova feedback broadens the density PDFs in all cases. However, ionisation feedback is more effective in triggering the high density part, which leads to triggered star formation, whereas the SN feedback also populates the low density part of the PDF. This shows that single SN explosions are mostly disruptive, and star formation is locally shut off. However, multiple SNe seem to be able to enhance the high density tail of the PDF, as can be clearly seen in case of FCB100 (blue dashed line), which has a dense tail at $10^3\;{\rm cm}^{-3} \lesssim n \lesssim 10^5\;{\rm cm}^{-3}$, which is missing in run FC.



\begin{figure*}
\includegraphics[width=88mm]{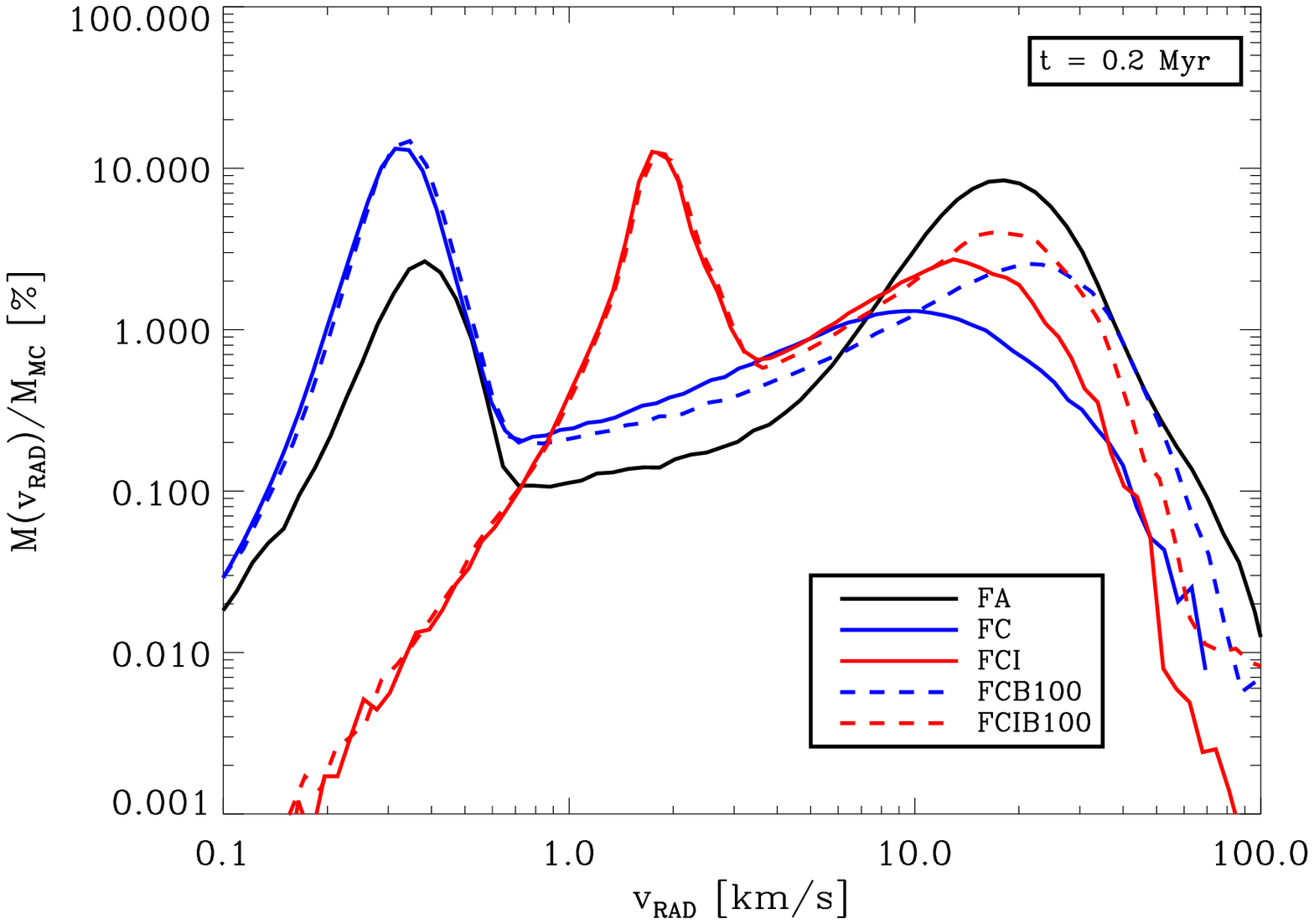} 
\includegraphics[width=88mm]{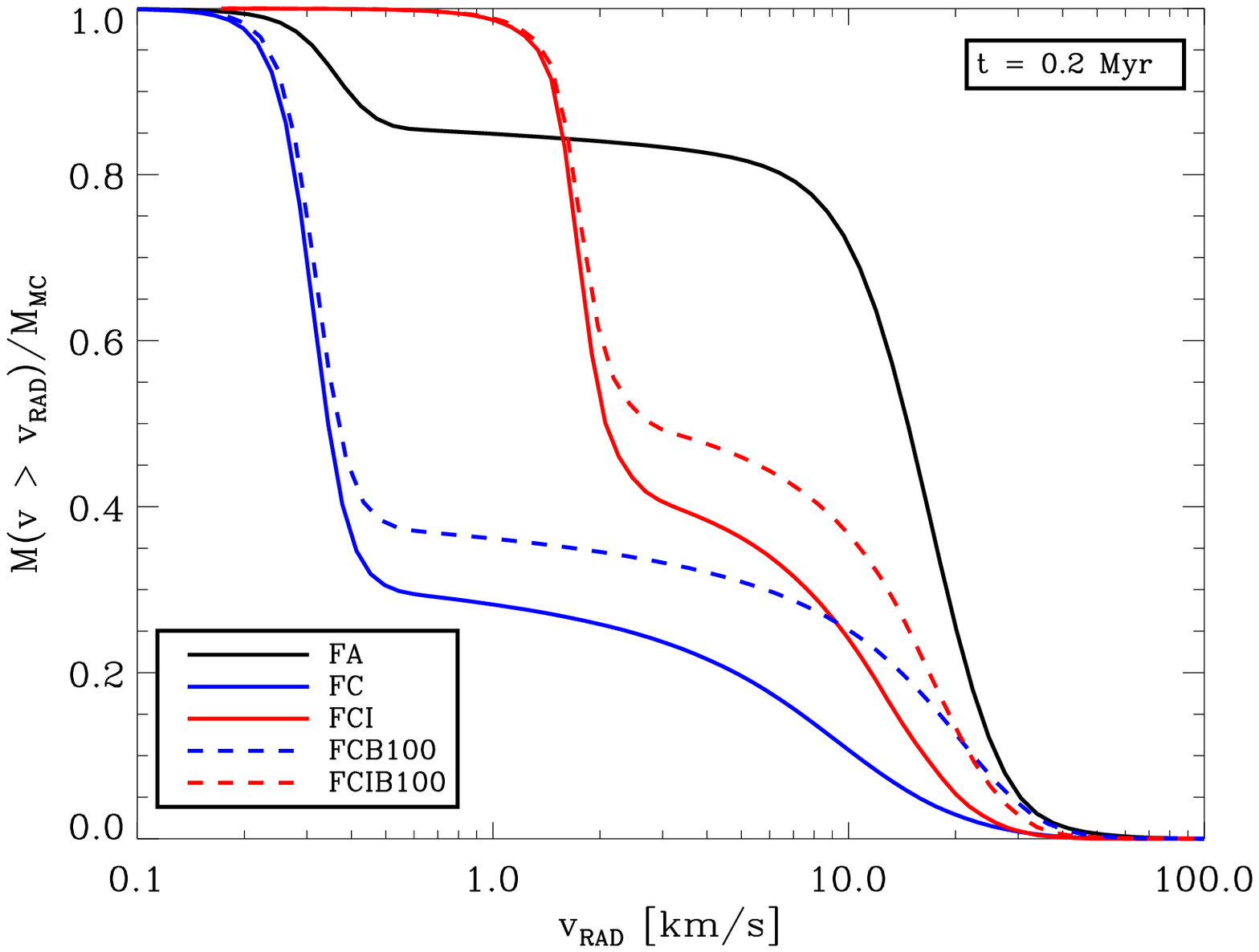} 
\caption{Radial velocity PDFs ({\it left panel}) and 100\% minus the cumulative distribution as a function of radial velocity ({\it right panel}) for runs FA, FC, FCI, and FCB100 at $t=0.2\,{\rm Myr}$. We note that the Mach number, i.e. the ratio of radial velocity to sound speed, is $M \approx 1$. This implies that only gas hotter than $\sim 10^4\,{\rm K}$ is leaving the cloud at velocities higher than $10\,{\rm km\,s^{-1}}$.  \label{FIG_VRAD}}
\end{figure*}
\begin{figure}
\includegraphics[width=88mm]{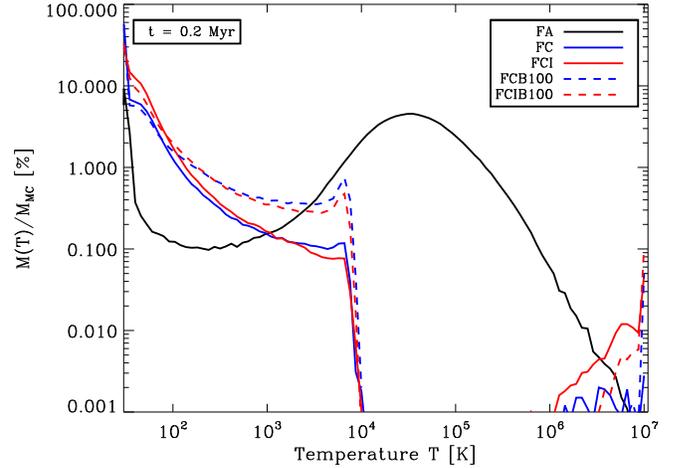} 
\caption{Temperature PDFs for runs FA, FC, FCI, and FCB100 at $t=0.2\,{\rm Myr}$. In the adiabatic simulation most of the gas resides in the thermally unstable regime and is cooled rapidly once radiative cooling is included. This leads to a bimodal distribution in all cases with radiative cooling. \label{FIG_TEMP}}
\end{figure}

\section{Energy and momentum input}\label{SEC_ENERGETICS}
In the following we discuss, which fraction of the energy and momentum input is transferred to the surrounding ISM. We measure the respective quantities at $t=0.2$ Myr. At this time most of the SN remnants are out of the ST stage and in the radiative stage, which means that their momentum does not increase anymore. In addition, no significant outflow has occurred, which might otherwise bias the results.

\subsection{Momentum input}\label{SEC_Pmom}
In Fig. \ref{FIG_EPMAX} we show the total cloud momentum measured at $t=0.2$ Myr after the first SN explosion, relative to the input momentum $P_\mathrm{SN}$, which is injected with the SN. For a single SN, $P_\mathrm{SN} = P_0 = 2.77 \times 10^4 \,{\rm M}_\odot\,{\rm km\, s}^{-1}$, and for all 'binary' runs which feature two explosions $P_\mathrm{SN} = 2\times P_0 = 5.54 \times 10^4 \,{\rm M}_\odot\,{\rm km\, s}^{-1}$ as listed in Table \ref{TABLE_1}.
Most SN remnants are past the ST phase at $t=0.2$ Myr. However, there are a few exceptions where the SN remnant is still in the ST phase at $t=0.2$ Myr. We mark this by an upward arrow, which indicates that the momentum will be able to increase further. 

{\sc Adiabatic runs:} The adiabatic runs, HA and FA, show that more than five times the input momentum can be transferred to the ISM in case radiative cooling is neglected. At $t=0.2$ Myr the shock front has almost reached the outer radius of the cloud in this case. As shown in Fig. \ref{FIG_VRAD} (black solid line), more than 70\% of the cloud mass has been accelerated to $v_{_{\rm RAD}}>\;10\;{\rm km/s}$ for run FA. Also, most of the gas is in the warm-hot, unstable regime with temperatures in between $10^4$ and $10^5$ K (see Fig. \ref{FIG_TEMP}). This gas is prone to experience quick radiative cooling processes and has cooled to $T\lesssim10^4$ K in all non-adiabatic runs. In total we measure $f_\mathrm{P} = P/P_\mathrm{SN} \sim 50$ for the adiabatic runs at $t=0.2$ Myr. A simple back-of-the-envelope calculation\footnote{If we assume that $\sim$30\% of the total energy stays in the form of kinetic energy (as in the ST phase, see section \ref{SEC_Einput}), we can compute the swept-up mass at the time when the shell has decelerated to 10 km/s and merges with the ISM. The swept-up mass ($\sim 3\times 10^5\;{\rm M_\odot} $) times 10 km/s then gives $f_\mathrm{P}\sim100$.} shows that the initial momentum should be enhanced by up to a factor of $f_\mathrm{P}\sim100$ in the adiabatic case until the remnant merges with the ISM. In Table 1 we list $f_\mathrm{P}$ for all runs.\\

{\sc Radiative Cooling:} From Fig. \ref{FIG_EPMAX} it is apparent that an adiabatic treatment of the SN remnant leads to a severe overestimation of the momentum input with respect to all other simulations we performed. All runs that include radiative cooling have $f_\mathrm{P}< 20$. For single explosions in media where the early stellar feedback has not been considered we have $f_\mathrm{P} < 10$ (run HC and FC). The gas in the warm-hot, thermally unstable regime ($1.2 \times 10^4\; {\rm K} \le T < 3 \times 10^5 $ K) cools quickly and leads to a bimodal temperature distribution (see Fig. \ref{FIG_TEMP}). \\

{\sc Ionising radiation:} If ionising radiation shapes the cloud before the SN explosion, $f_\mathrm{P}$ is about 50\% higher for both homogeneous (HC vs. HCI) and fractal clouds (FC vs. FCI). Although, the temperature structure appears to be similar for FC and FCI at $T\lesssim10^4$ K (see Fig. \ref{FIG_TEMP}, blue and red solid lines), slightly more hot gas ($T \gtrsim \;10^6$ K) remains in run FCI. The radial velocity distributions of FC and FCI (Fig. \ref{FIG_VRAD}) are quite different as almost all the gas is accelerated to $\sim 2\;{\rm km/s}$ in FCI, and $\sim$ 30\% of the gas is accelerated to $v_{_{\rm RAD}}>\;10\;{\rm km/s}$.\\

{\sc Lower ambient density:} When the ambient cloud density is reduced to $\bar{n}=10\;{\rm cm} ^{-3}$ (run FCN10) or $\bar{n}=1\;{\rm cm} ^{-3}$ (run FCN1), the momentum input is higher (by $\sim$ 50\%). FCN1 is still in the ST stage at $t=0.2$ Myr, and therefore the momentum will continue to increase. We have followed FCN1 until the end of the ST stage to make sure that $f_\mathrm{P} < 20$ in this case. \\

{\sc Binary explosions:} We also investigate the effect of a second explosion, which follows sooner or later after the first SN. In particular we have run FCB0, where two SNe are ignited at the same time; FCB5, where the second explosion takes place at $t=5$ kyr after the first one; FCB20, with the second SN after $t=20$ kyr, and FCB100, with the second explosion after $t=100$ kyr. We find that a second explosion which follows soon after the first one does not lead to a more efficient momentum input. It seems that the longer the time delay in between the two explosions, the more efficient is the second explosion, since the ST phase of the second SN lasts longer if the first SN shell is located at larger radii. Also the second explosion then hits a shell with larger surface area and mass.

When comparing the momentum input in runs that include the expansion of an H{\sc ii} region prior to the first SN and runs with a second explosion, it seems that the preparation of the cloud by ionising radiation allows for a higher momentum input (Fig. \ref{FIG_EPMAX}), while the second SN heats slightly more gas to $10^4$ K (Fig. \ref{FIG_TEMP}). From Fig. \ref{FIG_VRAD} we see that the second SN is responsible to shift $\sim$ 10\% of the gas to radial velocities larger than 25 km/s, while at lower velocities the distribution is mostly determined by the effect of the ionising radiation. \\

\subsection{Energy input}\label{SEC_Einput}

\begin{figure*}
\includegraphics[width=150mm]{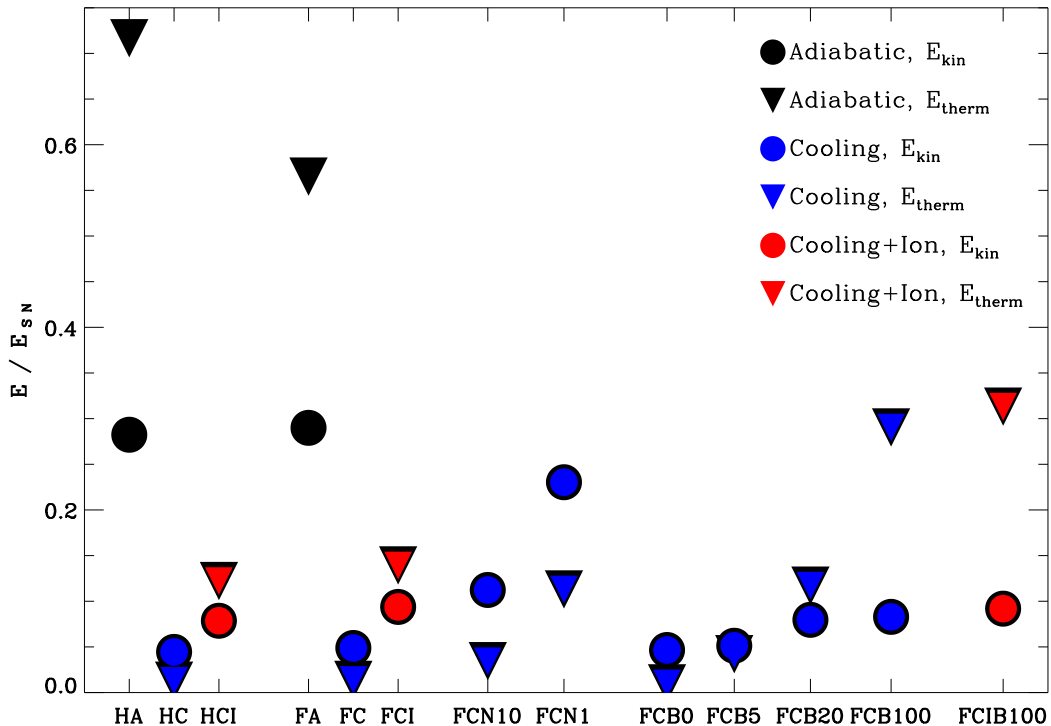} 
\caption{Gas thermal and kinetic energy input into the ISM at $t=0.2$
  Myr relative to the input ejecta energy of $E_\mathrm{SN} = 10^{51}\;{\rm
    erg}$ for a single SN and $E_\mathrm{SN} = 2\times10^{51}\;{\rm
    erg}$ for two SNe, respectively. The circles show kinetic energy
  retained by the ISM ($f_\mathrm{kin}$) and the triangles the thermal
  energy ($f_\mathrm{therm}$). Black symbols are used for adiabatic
  runs, blue symbols for runs with radiative cooling, and red symbols
  for runs with radiative cooling and ionisation. Again, the adiabatic
  cases (HA and FA) are still in the ST phase and largely
  over-estimate the energy budget with respect to runs that include
  radiative cooling. Ionisation prior to the SN (HCI, FCI, and
  FCIB100) store more thermal energy in $10^4$ K gas, which is
  thermally stable. This effect is also present in case of a second SN
  explosion within the cavity of the first one. However, only in lower
  density environments (FCN10 and FCN1), a significantly larger
  fraction of kinetic energy can be transferred to the ISM. The exact
  values of $f_\mathrm{kin}$ and $f_\mathrm{therm}$ are listed in
  Table \ref{TABLE_1}.   \label{FIG_EMAX}} 
\end{figure*}

The impact of the SN on a molecular cloud has typically been quantified in terms of the fraction of injected energy, which is transformed into thermal, $E_{_{\rm therm}}$, and kinetic energy, $E_{_{\rm kin}}$, of the surrounding ISM. The fraction of kinetic energy retained by the surrounding ISM is of particular interest since SNe are thought to be a major driver of interstellar gas turbulence. We term this quantity the {\it kinetic energy conversion efficiency}, $f_\mathrm{kin}=E_{_{\rm kin}}/E_{_{\rm SN}}$. The {\it thermal energy conversion efficiency} is $f_\mathrm{therm}=E_{_{\rm therm}}/E_{_{\rm SN}}$. In Table 1, we list $f_\mathrm{kin}$ and $f_\mathrm{therm}$ at $t=0.2$ Myr for all simulations.

The energy conversion efficiency is a long debated topic in the literature.
In the ST solution a constant ratio of thermal to kinetic energy of $E_{_{\rm therm}}/E_{_{\rm kin}} \sim 7/3$, or equivalently $f_\mathrm{kin}\sim 0.3$, arises. This result is independent of whether the SN energy was injected as thermal or kinetic energy. Therefore, the kinetic energy is always lower than the thermal energy of the gas as long as the system is not dominated by radiative cooling. With radiative cooling however, the thermal energy can be radiated away and lost. 

From early simulations, \citet{Cox1972} estimates that 10\% of the initial SN energy is retained as thermal and kinetic energy in the gas. \citet{Spitzer1968} show that 3\% of the initial SN energy is in gas motions when the motions have slowed to the mean velocity of interstellar clouds. \citet{Chevalier1974} obtain between 4\% and 8\%, depending on the ambient homogeneous density. Overall these results show that $f_\mathrm{kin}$ is generally low ($<$ 0.1) in homogeneous clouds.\\

In Fig. \ref{FIG_EMAX} we plot the kinetic and thermal energy left in the gas at $t=0.2$ Myr relative to the energy input of the SN(e), i.e. $E_\mathrm{SN} = E_0  = 10^{51}\,{\rm erg}$ and $E_\mathrm{SN} = 2 \times E_0  = 2\times 10^{51}\,{\rm erg}$, respectively (see Table \ref{TABLE_1}). Note that the potential energy as well as the kinetic and thermal energies induced by ionising radiation (runs HCI and FCI) are small and can be neglected for the purpose of this study. 

{\sc Adiabatic runs:}  In run HA we recover the ST solution (regime of $E_{_{\rm therm}}/E_{_{\rm kin}} \sim 7/3$), which is in agreement with the shock front evolution derived in section \ref{SEC_STRUCTURE}. Since the total energy in HA is supposedly constant, we may estimate the amount of numerical dissipation from this run. We find it to be in the percentage range and of the order of $\sim$ 3\% at the time where the shock front reaches $R_{_{\rm MC}}$.

{\sc Radiative cooling:} In run HC, radiative cooling is so strong that almost all of the injected energy is lost in a short time. At $t=0.2$ Myr, only $\sim$ 5\% of the total energy is retained, where $f_\mathrm{kin} \sim 0.04$ are in form of kinetic energy as well as 1\% of thermal energy ($f_\mathrm{therm} \sim 0.01$). In run HC an energetic outflow, which transports some of the SN energy into neighbouring regions, is completely absent. Thus, the SN impacts the cloud only locally and has radiated all its energy before reaching the edge of the cloud. At $t=1$ Myr the kinetic energy conversion efficiency goes down to $f_\mathrm{kin}(1\,{\rm Myr}) \approx 0$.

{\sc Ionising radiation:} As demonstrated in run HCI, the low-density,
warm environment of the SN, which is provided by an H{\sc ii} region
initially promotes the impact of the SN, as it effectively delays the
transition into the radiative stage. Therefore the gas has more time
to react to the SN explosion and a slightly higher fraction of the
injected energy is retained. At $t=0.2$ Myr we find
$f_\mathrm{therm}\sim 0.12$ for the thermal energy and $f_\mathrm{kin}
\sim 0.08$ for the kinetic energy. In addition, a small amount of
energy ($\sim$ 1\%) is carried away by outflowing gas (across the
molecular cloud boundary at $R_{_{\rm MC}}$). Nevertheless, after 1
Myr we find $f_\mathrm{kin} \approx 0.025$ and overall run HCI evolves
rather like HC than HA. 

{\sc Fractal clouds:} In all three fractal cloud runs (FA, FC, FCI) a major amount of energy is quickly transported out of the cloud. In particular, run FA looses mass and therefore energy in an energetic outflow. Thus, the fact that the retained energy appears to be smaller in run FA  than in HA is an artefact related to our analysis: Once gas outflow sets in, a significant fraction of energy is transported out of the cloud. However, this gas is flowing into vacuum and will therefore expand quickly and cool. Therefore, we do not consider gas that has left the cloud in the energy analysis. For runs FC and FCI, the energy evolution of the remaining gas is similar to the homogeneous cases.

{\sc Low density clouds:} The two runs with lower density (FCN10 and FCN1) show a higher kinetic energy fraction but small thermal energies. In particular FCN1 efficiently retains its kinetic energy and we find $f_\mathrm{kin} \sim 0.23$ and $f_\mathrm{therm} \sim 0.11$ at $t=0.2$ Myr.

{\sc Binary explosions:} Secondary SNe are only efficient if they are slightly delayed with respect to the first SN. Clearly, the retained energy increases with the time delay between the two explosions. We find that a delay of 20 kyr (run FCB20) gives a comparable result to run FCI, although the momentum input in FCI is still slightly higher.  Run FCIB100 with ionisation and a second explosion after 100 kyr results in the highest retained energies with respect to all other simulations except the adiabatic ones. However, we note that the retained energy is mostly thermal energy, and therefore could be quickly radiated away without being transferred to the surrounding ISM.


\begin{figure*}
\includegraphics[width=170mm]{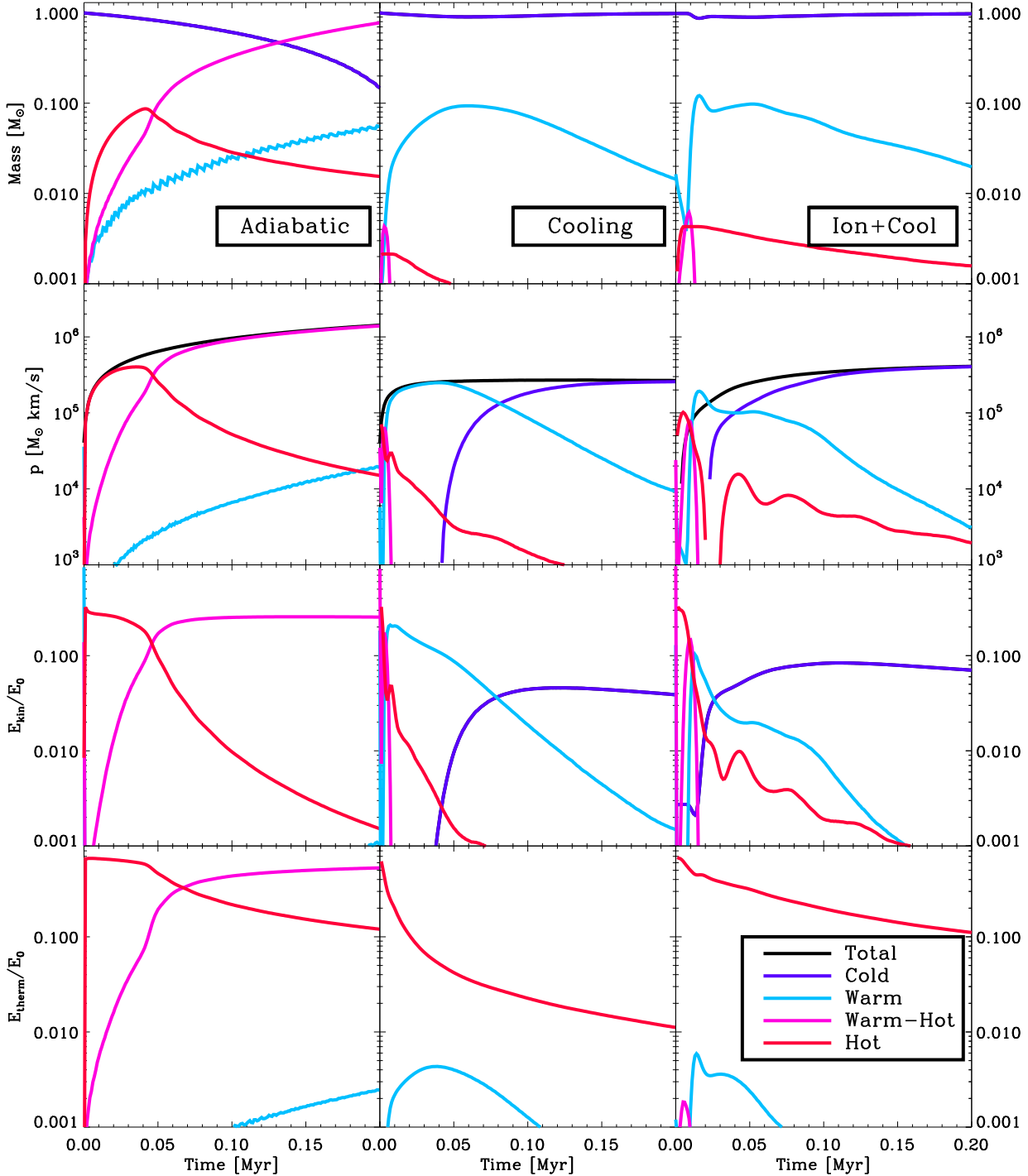} 
\caption{{\it Top to bottom:} Time evolution of cloud mass, momentum, kinetic, and thermal energy for runs HA (left column), HC (central column), and HCI (right column). We distinguish between four different phases, i.e. {\bf cold} ($T<300$ K; dark blue), {\bf warm} ($300\; {\rm K}\le T < 1.2 \times 10^4$ K; light blue), {\bf warm - hot} ($1.2 \times 10^4\; {\rm K} \le T < 3 \times 10^5 $ K; magenta), and {\bf hot} ($T \ge 3 \times 10^5$ K; red). The totals are shown in black in case they change as a function of time. For the adiabatic case, most of the mass is heated and resides in the thermally unstable regime ({\bf warm-hot} phase). If radiative cooling is included most of the mass is in {\bf cold} gas, with a slightly more massive {\bf hot} component in the run with cooling and ionisation. Similarly, the momentum and kinetic energy are mostly carried by the {\bf cold} gas. Only the thermal energy is carried by the {\bf hot} gas. Since there is more {\bf hot} gas in run HCI than in run HC, the thermal energy is higher in this case. The evolution of the thermal energy carried by the {\bf hot} gas is comparable in runs HA and HCI. \label{FIG_ALL_UNI}}
\end{figure*}

\begin{figure*}
\includegraphics[width=170mm]{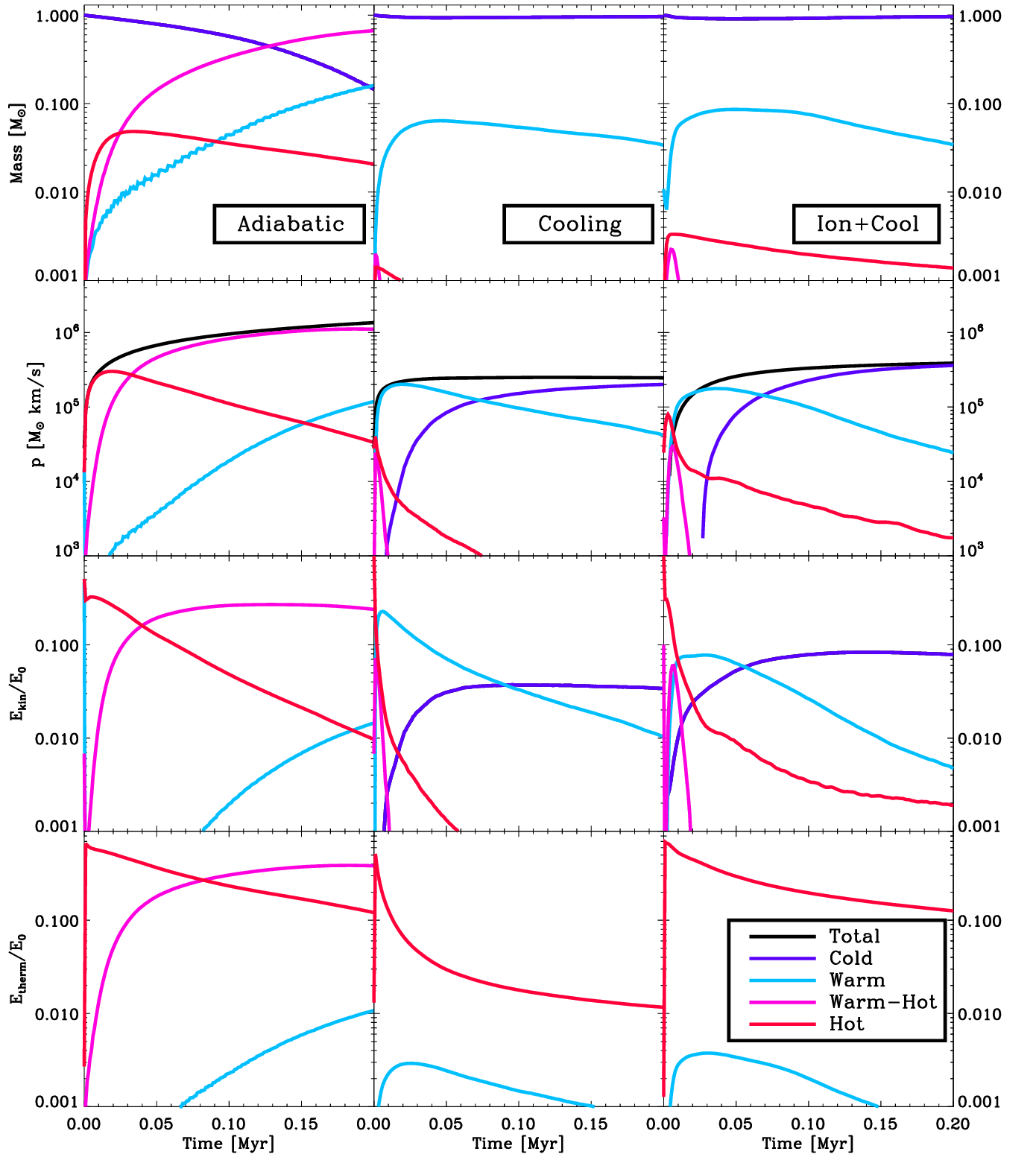} 
\caption{Same as figure \ref{FIG_ALL_UNI} but for runs FA (left column), FC (central column), and FCI (right column). The cloud structure has a minor impact on the evolution of mass budget, momentum, kinetic and thermal energies. The only obvious difference with respect to Fig. \ref{FIG_ALL_UNI} is a more abundant {\bf warm} component in both runs with radiative cooling (FC and FCI), which leads to a slightly higher input of kinetic energy than in HC and HCI.}\label{FIG_ALL_F}
\vspace{2cm}
\end{figure*}

\section{Mass, momentum, and energy evolution}\label{SEC_OUTFLOW}
To investigate the SN efficiency we distinguish between four different temperature regimes 
\begin{itemize}
\item {\bf Cold}: $T<300\;{\rm K}$; the stable cold gas phase.
\item {\bf Warm}: $300\; {\rm K}\le T < 1.2 \times 10^4$ K; the warm phase, consisting of warm atomic or ionised gas. It should be noted that, for runs HI and FI, the gas heated to 10,000 K within the HII region contributes only little to the total mass in the cloud.
\item {\bf Warm-Hot}: $1.2 \times 10^4\; {\rm K} \le T < 3 \times 10^5 $ K; the unstable hot phase.
\item {\bf Hot}: $T \ge 3 \times 10^5$ K; gas in the 'stable' hot phase. This gas has a sound speed of more than $40\;{\rm km/s}$.\\
\end{itemize}

In Fig. \ref{FIG_ALL_UNI} and \ref{FIG_ALL_F} we show the following quantities, split up by temperature regime:
\begin{itemize}
\item[1.] The time evolution of the enclosed gas mass ($R < R_{_{\rm MC}} = 16$ pc, top row).
\item[2.] The momentum evolution of the enclosed gas (2$^{\rm nd}$ row).
\item[3.] The normalised kinetic energy (3$^{\rm rd}$ row), and
\item[4.] the thermal energy (4$^{\rm th}$ row).
\end{itemize}
Fig. \ref{FIG_ALL_UNI} depicts the homogeneous cases. From left to right we show the adiabatic run (HA), the run with radiative cooling (HC), and radiative cooling and initial H{\sc ii} region (HCI). In Fig. \ref{FIG_ALL_F} we plot the fractal cases with a single SN event (from left to right: FA, FC, FCI). \\

In the adiabatic cases, most of the gas is in the warm-hot regime, which is thermally unstable and therefore not populated in the case that radiative cooling is included. Thus, for all runs including radiative cooling, most of the gas remains cold. Since the mean radial velocities are moderate, the phase which dominates the mass budget also carries most of the momentum and kinetic energy. However, most of the thermal energy is carried by the hot phase of course, which contributes a negligible mass fraction.

Runs with an initial H{\sc ii} region feature more warm and hot gas than pristine clouds. Therefore, they may retain more of the SN energy and the transferred momentum is increased.

The evolution of the remnant is more complex for structured clouds (Fig. \ref{FIG_ALL_F}), since the cloud substructure changes the progression of the SN as the remnant encounters different column densities along different radial directions.
Nevertheless, a cloud with sub-structure does not follow a dramatically different mass or momentum evolution. The kinetic energy conversion efficiency is slightly higher in structured media because the density PDF is broader and so some more warm gas ($10^4$ K) may survive, but we are speaking about changes of the order of $\lesssim$1\%. In dense media, where most of the energy is lost, it is also lost for gas with a sub-structure, as long as the sub-structure can be described with a typical fractal dimension.

\section{Conclusions}\label{Conclusions}
To investigate the impact of supernova explosions on the evolution of
cold structured and partially ionized molecular clouds, we perform
high resolution SPH simulations of single Supernova explosions in
molecular clouds of homogeneous or fractal structure with the SPH code
SEREN. All clouds have a fixed mass of $10^5 \;{\rm M}_\odot$. With a
radius of 16 pc, this results in a typical mean number density of
$\bar{n}= 100 \;{\rm cm}^{-3}$. We 
also explore $\bar{n} = 10 \;{\rm cm}^{-3}\;{\rm and}\; 1 \;{\rm
  cm}^{-3}$. In addition to a simple adiabatic equation of state which
serves as a reference case, we compare the effect of radiative cooling
and an initial H{\sc ii} region which has self-consistently evolved by
following the ionizing radiation of the progenitor star within the
molecular cloud. 

For the homogenous adiabatic case we recover the classic Sedov-Taylor
solution. The gas is rapidly heated and the cloud is dispersed within
$<$1 Myr. However, in this case, most of the gas remains in the
warm-hot ($1.2\times 10^4\;{\rm K} \le T < 3 \times 10^5 \; {\rm K}$),
thermally unstable phase, and cools on very short time scales if
cooling is included. We confirm that the momentum and energy imparted
to the ISM are severely over-estimated if radiative cooling is not
included. 

In the presence of radiative cooling, the cloud is not dispersed. After a short initial Sedov-Taylor phase the thermally unstable gas cools rapidly and after $<$ 8,000 years after the explosion the cloud enters the pressure dominated snowplough phase. Almost all the gas is now warm ($300\;{\rm K}\; \le T < 1.2\times 10^4\;{\rm K}$) or cold ($T < 300\;{\rm K}$). If the cloud was pre-ionized by the progenitor star, the SN shock wave initially moves faster, radiative losses are slightly reduced and the momentum input is increased by a factor of $\sim$ 50\% with respect to the cases without an initial H{\sc ii} region. 

For SN explosions in fractal clouds, i.e. clouds with a density
substructure, the spherical symmetry of the problem is broken and no
global analytic solution can be easily applied. Depending on the
radial direction, the shock will encounter columns with varying
densities. Therefore, a fraction of the gas remains in the hot phase
($T \ge 3 \times 10^5  \;{\rm K}$) and may flow out of the cloud at
high radial velocities. Typically, the outflow is transsonic. This
implies that keeping the gas hot is a necessary condition for the
generation of fast outflows. However, the momentum and kinetic energy
input as well as their distribution onto the different temperature
regimes are almost indistinguishable for homogeneous and fractal
clouds. In summary, the sub-structure of the molecular clouds mostly
affects the outflow of gas, which is initiated by the SN
explosion. Since the thermal energy is mostly carried by the low mass,
hot component a high mass resolution is necessary in order to follow
high velocity, low mass outflows from the cloud. 

In the most realistic case of a SN exploding in a previously ionised,
structured cloud (run FCI), the H{\sc ii} region does not only heat
the gas in the vicinity of the SN explosion but also re-shapes the
cloud's density structure, i.e. it broadens the density PDF of the
molecular cloud prior to the explosion. The impact of the SN is
enhanced in some specific areas where the column density has been
decreased. In these areas, gas may quickly flow out of the cloud
(already at $t \gtrsim 0.1$ Myr) and provides the cloud environment
with hot gas that flows with $v_{_{\rm RAD}} \sim$ 80 km\/s. However,
the progress of the SN remnant is quickly stopped in areas where it
hits a dense shell or pillar that has been swept-up by the ionising
radiation. Therefore, ionising radiation does overall {\bf not} 
substantially enhance the impact of a SN explosion on a dense
molecular as it drives the ISM into inert dense shells and cold
clumps, a process which is unresolved in galaxy scale simulations. SNe
which explode within H{\sc ii} regions are only able to deposit $\sim$
50\% more momentum into the surrounding ISM (compared to a factor of
$\sim$ 50 higher momentum in adiabatic simulations). However, the
coulds structured by ionising radiation might significantly enhance
the escape fractions of UV photons which in turn might affect the ISM
on larger scales (see e.g. \citep{2014MNRAS.437.2882K}). 

We also confirm that the SN impact is increased in lower density
environments. In these, the SNe are able to transfer more kinetic
energy than thermal energy to the surrounding ISM. 
Furthermore, subsequent SNe seem to be able to initiate super-bubbles
{\it only} if they are well timed, i.e. if the second explosion
happens in the snowplough phase of the first one. Due to the low
coupling efficiencies our results support previous conclusions that
supernovae might only drive a wind if a significant fraction explodes
in low-density environments or if they are supported on larger scales
by processes other than ionising radiation. Single scattering seems
insufficient (e.g. \citet{2014MNRAS.439.2990S} and references therein)
but infrared radiation pressure
\citep{2011MNRAS.417..950H,2013ApJ...770...25A,2013MNRAS.434.2329K,2014arXiv1403.1874D} 
and/or cosmic rays
\citep{2012MNRAS.423.2374U,2013ApJ...777L..38H,2013ApJ...777L..16B,2014MNRAS.437.3312S}
are plausible options.


\section*{Acknowledgments}
SW and TN acknowledge J.P. Ostriker for interesting discussions on the subject.
SW acknowledges David Hubber and Thomas Bisbas for helpful comments
and suggestions. SW and TN acknowledge support from the DFG priority
programme 1573. The simulations were performed on SuperMUC at the
Leibniz-Rechenzentrum Garching and on {\sc ODIN} at the Rechenzentrum
Garching. The column density and temperature plots were made using the
SPLASH visualization code \citep{Price2007}. 

\bibliographystyle{mn2e}
\bibliography{references}

\clearpage

\label{lastpage}
\end{document}